\documentclass{jps-cp}
\usepackage{txfonts} %Please comment out this line unless the txfonts package is availabe in your LaTeX system.

\title{Effective Spin-1/2 Moments on a Yb$^{3+}$ Triangular Lattice: an ESR Study}

\author{J\"org \textsc{Sichelschmidt}$^{1}$, Burkhard \textsc{Schmidt}$^{1}$, Philipp \textsc{Schlender}$^{2}$, Seunghyun \textsc{Khim}$^{1}$, Thomas \textsc{Doert}$^{2}$, and Michael \textsc{Baenitz}$^{1}$}

\inst{
$^{1}$ Max Planck Institute for Chemical Physics of Solids, Dresden, Germany\\
$^{2}$ Faculty of Chemistry and Food Chemistry, TU Dresden, Dresden, Germany\\
}

\email{Sichelschmidt@cpfs.mpg.de}

\recdate{August 27, 2019}

\abst{We investigated the spin dynamics by electron spin resonance (ESR) of the Yb-based, effective spin-1/2 delafossites NaYbO$_{2}$, AgYbO$_{2}$, LiYbS$_{2}$,  NaYbS$_{2}$, and NaYbSe$_{2}$ which all show an absence of magnetic order down to lowest reachable temperatures and thus are prime candidates to host a quantum spin-liquid ground state in the vicinity of long range magnetic order. Clearly resolved ESR spectra allow to obtain well-defined $g$ values which are determined by the crystal field of the distorted octahedral surrounding of the Yb-ions in trigonal symmetry. This local crystal field information provides important input to characterize the effective $S = 1/2$ Kramers doublet as well as the anisotropic exchange coupling between the Yb ions which is crucial for the nature of the groundstate.   
The ESR linewidth $\Delta B$ is characterised by the spin dynamics and is mainly determined by the anisotropic exchange coupling. We discuss and compare $\Delta B$ of the above mentioned delafossites focussing on the low temperature behaviour which is dominated by the growing influence of spin correlations. }

\kword{spin dynamics, triangular lattice, quantum spin liquid, electron spin resonance}

\begin{document}
\maketitle

\section{Introduction}

Recently it was shown that Yb-based delafossite systems are an ideal platform to study spin-orbit entangled frustration effects on a perfect triangular lattice \cite{baenitz18a,liu18a,ranjith19a}. They are possible candidates to host a quantum spin-liquid (QSL) ground state \cite{zhou17a} which is characterized by persistent magnetic fluctuations down to zero temperature accompanied by an absence of long range magnetic order.
Especially for systems with strong spin-orbit interaction theory predicts new exotic spin states (including different QSLs) with unconventional excitations \cite{maksimov19a}.
%Electron Spin Resonance (ESR) was shown to be an informative spectroscopic probe of spinons, the elementary excitations in a QSL \cite{luo18a}. 
Electron Spin Resonance (ESR) was used to characterize the spin dynamics of potential QSL systems with Yb$^{3+}$ spins on a triangular lattice \cite{li15a,sichelschmidt19a} where anisotropic exchange interactions between the spins have a major impact on the ESR linewidth. A clear and well-resolved Yb$^{3+}$ spin resonance was recently reported for the delafossite system NaYbS$_{2}$ providing local information on the Yb$^{3+}$ $g$-factor and spin dynamics \cite{sichelschmidt19a}. For theoretical treatments for instance on the anisotropic exchange of triangular-lattice Yb magnets \cite{rau18a} ESR-obtained $g$ factors are important parameters to compare with. 
Here we present ESR results on various Yb-based delafossite systems differing in their anisotropic exchange and crystal electric field splitting of the Yb Kramers doublets.

\section{Experimental}

\subsection{Electron Spin Resonance}

We used a standard continuous-wave ESR technique at X-band microwave frequencies ($\nu $=9.4 GHz). The resonance signal was recorded in the field-derivative d$P$/d$B$ of the absorbed power $P$ of a transversal magnetic microwave field $b_{\rm mw}$. The sample temperature was set with a helium-flow cryostat allowing for temperatures between 2.7 and 300~K.

The obtained spectra were fitted by a Lorentzian line shape yielding the parameters linewidth $\Delta B$ which is a measure of the spin-probe relaxation rate, and resonance field $B_{\rm res}$ which is determined by the effective $g$-factor $[g=h\nu/(\mu_{\rm B}B_{\rm res})]$ and internal fields. The spectra of the powder samples were fitted by directional averaging the Lorentzians with uniaxial anisotropy. 
The ESR intensity $I_{\rm ESR}$ is determined by the static spin-probe susceptibility $\chi_{\rm ESR}$ along $b_{\rm mw}$. Thus, $I_{\rm ESR}$ provides a direct microscopic probe of the sample magnetization. Details on the determination of $I_{\rm ESR}$ from the spectra are given in Ref.\cite{sichelschmidt19a}.
% is related to the integrated ESR absorption $I_{A}$ as $I_{\rm ESR}=I_{A}g$ where $g$ is the g-value component along $B$ \cite{gruner10a}. $I_{A}$ was calculated using the line amplitude and line width as reported earlier \cite{wykhoff07b}. 

\subsection{Sample preparation and characterisation}

We investigated single crystals of NaYbSe$_{2}$ as well as polycrystalline material of NaYbO$_{2}$, AgYbO$_{2}$, LiYbS$_{2}$,  NaYbS$_{2}$. All these compounds have a delafossite structure (space group $R\bar{3}m$) with Yb$^{3+}$ occupying a single crystallographic site having $\bar{3}m$ symmetry. The Ag-ion in AgYbO$_{2}$ is linearly coordinated with O, whereas Li- and Na-ions are octahedrally coordinated with the chalcogenides. 
%
%Single crystals of NaYbS$_{2}$ (reactive gas synthesis) and 
Polycrystalline powders of NaYbS$_{2}$ and LiYbS$_{2}$ (ampule-based synthesis) were prepared according to the procedures reported in Ref. \cite{baenitz18a}. NaYbSe$_{2}$ single crystals were prepared following the method of Lissner and Schleid \cite{schleid93a}.
Polycrystalline NaYbO$_{2}$ and AgYbO$_{2}$ were synthesized by a solid state reaction and a cation-exchange reaction, respectively \cite{saito08a}.
Magnetic and thermal properties were recently reported in Refs. \cite{ranjith19a,liu18a,ranjith19b}

\section{Results}

\begin{figure}[b]
\begin{center}
\includegraphics[width=0.7\columnwidth]{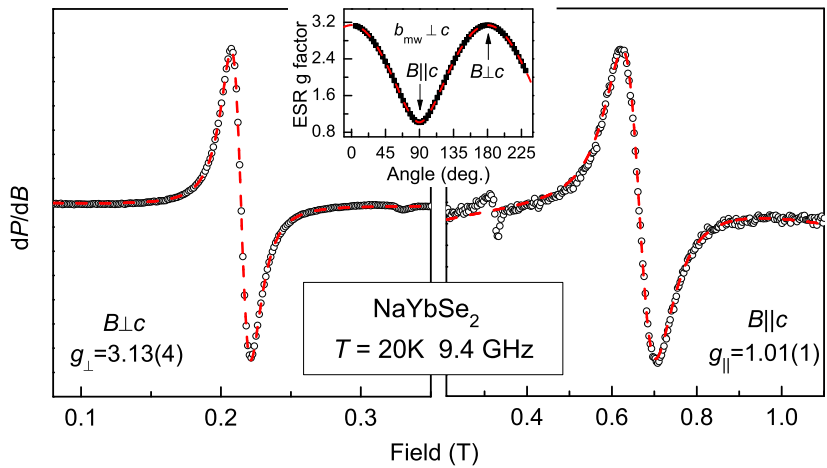}
%\vspace{4cm}
\caption{Typical ESR spectra of NaYbSe$_{2}$ for two crystal orientations. Dashed lines denote Lorentzian shapes with $g$ factors as indicated. The structure near 0.33~T is a background feature. Inset: $g$ factor dependence on the angle $\Theta$ between the external field $B$ and the crystallographic $c$-axis with the microwave field $b_{\rm mw}\perp c$. Dashed line indicates $g(\Theta) = \sqrt{g_\|^2\cos^2\Theta + g_\bot^2\sin^2\Theta}$ with $g_\bot$ and $g_\|$ given in the main frames.}
\label{FigNYSe1}
\end{center}
\end{figure}

We obtained and analyzed the ESR data of the above mentioned compounds in the same way as previously reported for NaYbS$_{2}$ \cite{sichelschmidt19a}. All spectra are well-resolved and reveal 
a common characteristics for a $J=7/2$ state of Yb$^{3+}$ in an crystal electric field environment with a $R\bar{3}m$ space group symmetry. 

\subsection{NaYbSe$_{2}$ single crystal}
Similar to NaYbS$_{2}$ \cite{sichelschmidt19a}, the ESR spectra of NaYbSe$_{2}$, shown in Fig. \ref{FigNYSe1}, have very clear amplitudes and relatively narrow linewidths as compared to the ESR spectra of YbMgGaO$_{4}$ \cite{li15a}. The magneto-crystalline anisotropy of Yb$^{3+}$ in an uniaxial crystalline-electric field yields the large anisotropy of the $g$-factor illustrated in the inset. 

\begin{figure}[tbh]
\begin{center}
\includegraphics[width=0.7\columnwidth]{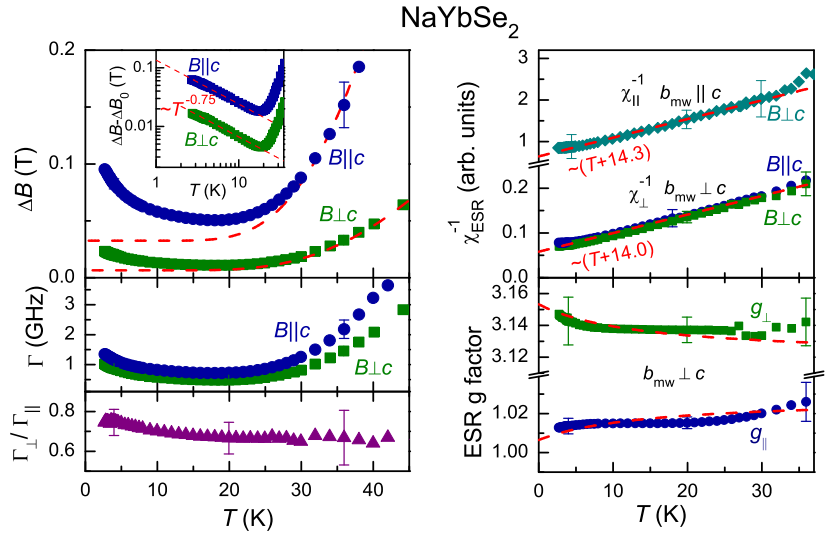}
%\vspace{4cm}
\caption{Temperature dependencies of ESR parameters of NaYbSe$_{2}$ for two orientations of the external field $B$ to the $c$-axis. Left: ESR linewidth $\Delta B$, relaxation rate $\Gamma=\nu\Delta B/B_{\rm res}$ and ratio of $\Gamma$ along the two directions of the field. Dashed lines describe $\Delta B(T)$ towards higher temperatures as a relaxation via the first excited crystalline electric field level of Yb$^{3+}$ at $\Delta=160\pm30$~K. Inset:  Linewidth without a residual contribution ($\Delta B_{0}$ from the dashed lines in the $\Delta B$-frame). Solid lines suggest a power law behavior as indicated. Right: ESR intensity $\chi_{\rm ESR}$ and $g$ factor for the external field $B$ and the microwave field $b_{\rm mw}$ aligned to the $c$-axis as indicated. 
Dashed lines denote Curie-Weiss fits for the intensity and fits of $g_\|(T)$ and $g_\bot(T)$ using the equations given in Ref. \cite{sichelschmidt19a} 
}
\label{FigNYSe2}
\end{center}
\end{figure}

Fig. \ref{FigNYSe2} demonstrates the temperature dependence of the ESR parameters being again very similar to NaYbS$_{2}$ \cite{sichelschmidt19a} . The linewidth $\Delta B$ and the relaxation rates $\Gamma$ show characteristic low- and high temperature behaviors: towards low temperatures a power law increase $\Delta B(T) \propto 1/T^{3/4}$ and towards high temperatures an Orbach process $\Delta B \propto 1/\exp(\Delta/k_{\rm B}T)-1$ which includes population of the first excited crystalline-electric field doublet at an energy $\Delta$ above the ground state \cite{sichelschmidt19a}. Within data accuracy, no clear temperature dependence of the $\Gamma$ anisotropy could be resolved. Similar to NaYbS$_{2}$ \cite{sichelschmidt19a} we obtained a small anisotropy of the Weiss temperatures from the temperature dependencies of ESR intensity and $g$-factor (Fig. \ref{FigNYSe2} right frame). For $g_\|(T)$ and $g_\bot(T)$ we used the equations given in Ref. \cite{sichelschmidt19a} with the parameters 
$g_{\bot}^0 \approx 3.12$ and $g_{\|}^0 \approx 1.03$, in good agreement with the values describing the anisotropy (Fig.\ref{FigNYSe1}).
 
\subsection{Polycrystalline NaYbO$_{2}$, AgYbO$_{2}$, LiYbS$_{2}$,  and NaYbS$_{2}$}

\begin{figure}[hbt]
\begin{center}
\begin{minipage}[c]{0.45\textwidth}
\includegraphics[width=1\columnwidth]{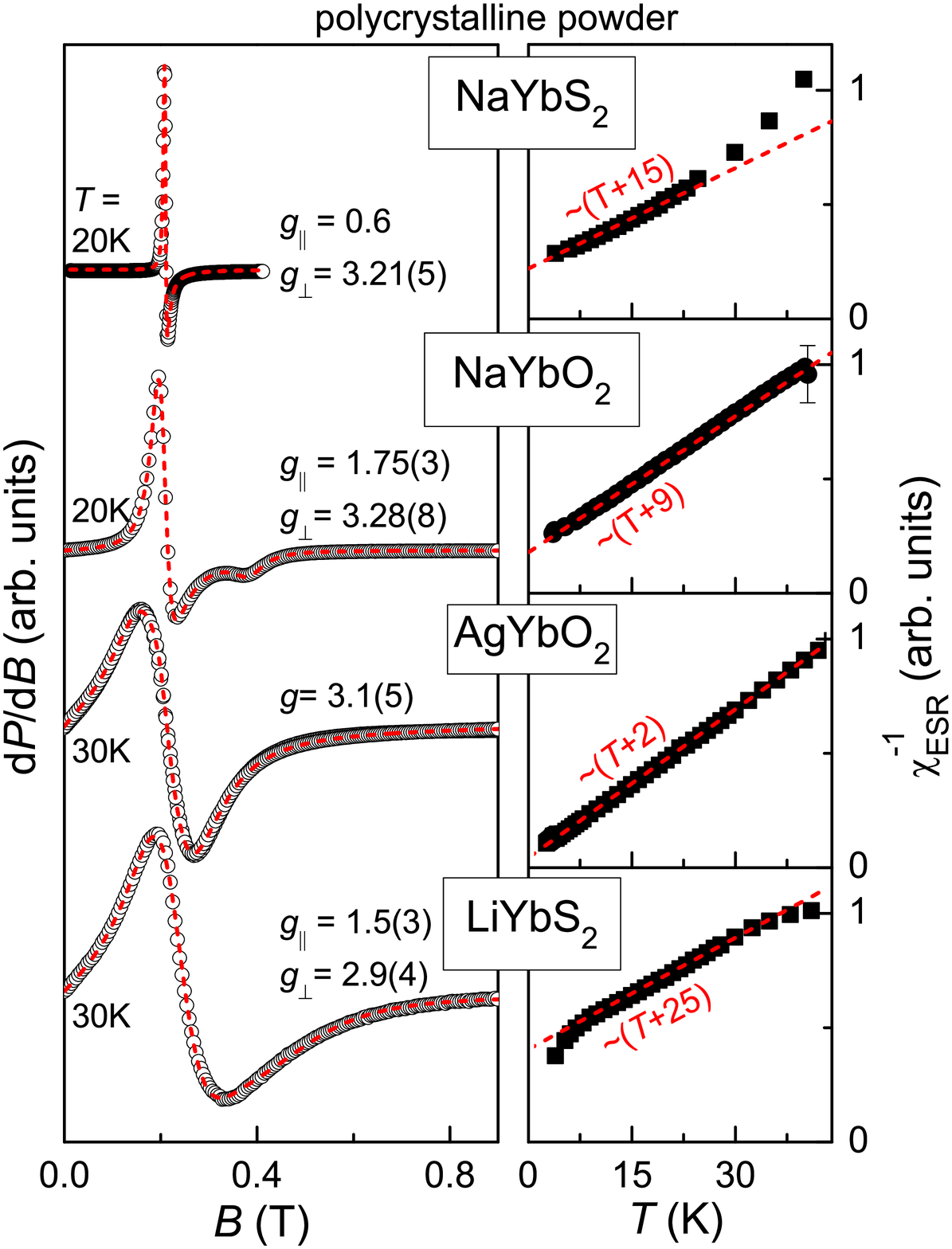}
%\vspace{4cm}
\caption{Left frame: Typical ESR spectra of the investigated polycrystalline powder samples at indicated temperatures. Dashed lines indicate a powder-averaged Lorentzian lineshape (except for AgYbO$_{2}$: single Lorentzian line). Right frame: inverse ESR intensity with dashed lines corresponding to a Curie-Weiss behavior as indicated.}
\label{FigPoly1}
\end{minipage}
\hfill
\begin{minipage}[r]{0.45\textwidth}
\includegraphics[width=1\columnwidth]{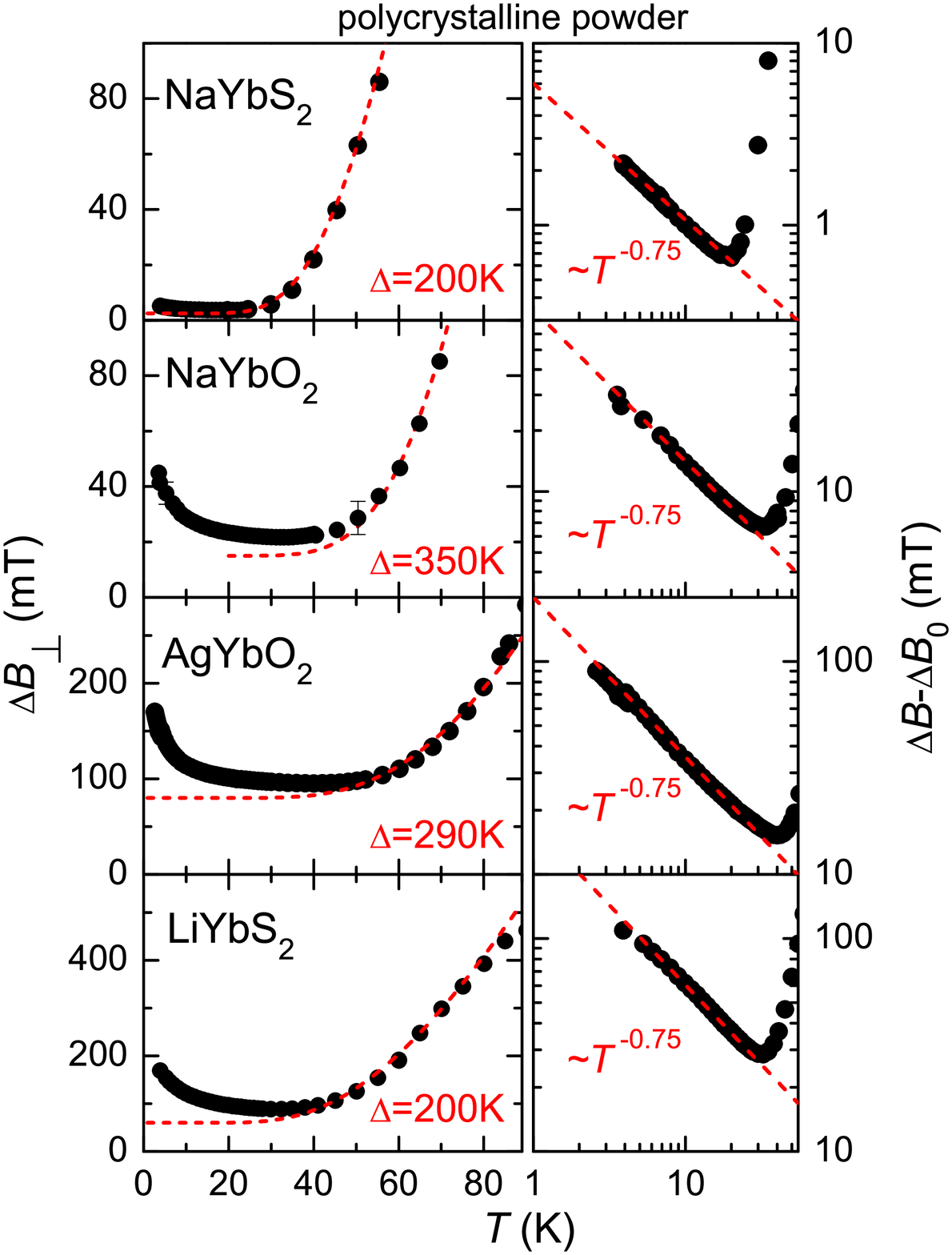}
%\vspace{4cm}
\caption{Left frame: Temperature dependence of ESR linewidth $\Delta B$ of the investigated polycrystalline powder samples. Dashed lines describe $\Delta B(T)$ towards higher temperatures as a relaxation via the first excited crystalline electric field level of Yb$^{3+}$ at $\Delta$~K. Right frame:  Linewidth without a residual contribution ($\Delta B_{0}$ from the dashed lines in the left frame). Dashed lines suggest a power law behavior as indicated.}
\label{FigPoly2}
\end{minipage}
\end{center}
\end{figure}

As shown in Fig. \ref{FigPoly1} the investigated polycrystalline samples all show well-defined powder-shaped Lorentzian ESR spectra with no dependence on the orientation of the field - as expected for arbitrarily oriented microcrystallites. The ESR intensity follows a Curie-Weiss law fairly well below about $T=30$~K as suggested by the red dashed lines in the right frame of Fig. \ref{FigPoly1}.

Figure \ref{FigPoly2} compiles the temperature dependence of the linewidth for the polycrystalline samples. Again, as indicated by the dashed red lines, at high temperatures the first excited crystalline electric field level located at $\Delta$ dominates (left frame). For NaYbO$_{2}$ the ESR-obtained $\Delta=350\pm30$~K roughly agrees with the value $\Delta\simeq400$~K obtained from neutron scattering \cite{ding19a}. Towards low temperatures the linewidth increases with a power law behavior $\Delta B(T) \propto 1/T^{3/4}$ (right frame).   

%\begin{figure}[t]
%\begin{center}
%\includegraphics[width=0.5\columnwidth]{FigdHPowder2.pdf}
%%\vspace{4cm}
%\caption{Left frame: Temperature dependence of ESR linewidth $\Delta B$ of the investigated polycrystalline powder samples. Dashed lines describe $\Delta B(T)$ towards higher temperatures as a relaxation via the first excited crystalline electric field level of Yb$^{3+}$ at $\Delta$~K. Right frame:  Linewidth without a residual contribution ($\Delta B_{0}$ from the dashed lines in the left frame). Dashed lines suggest a power law behavior as indicated.}
%\label{FigPoly2}
%\end{center}
%\end{figure}

\section{Discussion and Summary}
All investigated samples show a quite comparable characteristics of their Yb$^{3+}$ spin resonance, noteworthy a large g factor anisotropy, a clear influence of the first excited crystal-field split level on the linewidth and (at least for the investigated single crystals) a weak anisotropy of the local susceptibility obtained from the intensity. Table \ref{t1} compiles the main parameters extracted from the ESR data presented in the sections above. 

\begin{table}[tb]
\caption{ESR parameters with two uniaxial components: $g$-factor obtained from fitting the spectra by a Lorentzian shape, Weiss temperature $\theta$ obtained from the ESR intensity $\chi^{-1}_{\rm ESR}\propto (T+\theta)$, and first excited crystalline electric field level $\Delta$ ($\pm 30$~K) obtained from the high temperature behavior of the linewidth.}
\label{t1}
\begin{tabular}{ccccccccc}
\hline
compound & & $g_\|$ & $g_\bot$ & $\theta_\|$(K) & $\theta$ (K) & $\theta_\bot$ (K) & $\Delta$(K) & remark \\
\hline
NaYbSe$_{2}$ & crystal & 1.01(1) & 3.13(4) & 14.3 &-& 14.0 & 160 & \\
NaYbS$_{2}$  & crystal & 0.57(3) & 3.19(5) & 15.2 &-& 14.8 & 198 & from \cite{sichelschmidt19a} \\
NaYbS$_{2}$ & polycrystal & 0.6 & 3.21(5) &-& 15 &-& 200 & $g_\|=0.6$ fixed param. \\
NaYbO$_{2}$ & polycrystal & 1.75(3) & 3.28(8) &-& 9 &-& 350 & \\
AgYbO$_{2}$ & polycrystal &-&3.1(5)&-& 2 &-& 290 & \\
LiYbS$_{2}$ & polycrystal & 1.5(3) & 2.9(4) &-& 25 &-& 200 & \\
\hline
\end{tabular}
\end{table}

The anisotropy in the Weiss temperatures obtained from the single crystal ESR intensity of about 0.4~K is rather weak when compared with magnetization results \cite{baenitz18a,ranjith19b}. 
%(the reason may be due to different regions of linear temperature behavior in $\chi^{-1}(T)$). 
Note, however, that the exchange anisotropy provides a considerable contribution to the linewidth. As was estimated for NaYbS$_{2}$ \cite{sichelschmidt19a} hyperfine and dipolar broadening yield less than 1~mT whereas the smallest observed linewidth amounts to 3.6~mT (in polycrystalline NaYbS$_{2}$).  

Regarding a putative presence of a QSL ground state the low-temperature behavior of the linewidth is worth to be considered, keeping in mind that for none of the investigated samples magnetic order was observed. 
For NaYbO$_{2}$ the onset of the low-temperature increase of $\Delta B$ occurs in a similar temperature region where also the muon spin relaxation rate increases, tracking the onset of correlations between Yb$^{3+}$ pseudospins \cite{ding19a}. 
Below $T=20-30$~K down to the lowest accessible temperatures (2.7~K) a power law increase $\Delta B(T) \propto 1/T^{3/4}$ seems reasonable for all investigated samples (see Figs.~\ref{FigNYSe2}, \ref{FigPoly2}). Such power law behavior indicates a suppression of exchange narrowing by classical critical fluctuations of a 3D order parameter \cite{forster13a}. 

\section{Acknowledgements}
T.D. and P.S. thank the Deutsche Forschungsgemeinschaft for financial support within the CRC1143 framework (project B03). We acknowledge Dieter Ehlers for developing a powerful routine for fitting the ESR spectra, see {http://myweb.rz.uni-augsburg.de/\~{}ehlersdi/spektrolyst/} and Stefan Koch for assisting the ESR measurements.

\end{document}